\documentclass[acmlarge, nonacm]{acmart}

\AtBeginDocument{%
  }

\setcopyright{none}

\graphicspath{{./images/}{./figures/}}

\usepackage{booktabs}
\usepackage{subfig}

\begin{document}

\title{Systematizing Blockchain Research Themes and Design Patterns:\\
Insights from the University Blockchain Research Initiative (UBRI)}

\author{Chien-Chih Chen}
\email{j2255che@uwaterloo.ca}
\affiliation{%
  \institution{University of Waterloo}
  \city{Waterloo}
  \state{Ontario}
  \country{Canada}
}

\author{Yitian Wang}
\email{tammy.wang.20@ucl.ac.uk}
\affiliation{%
  \institution{University College London}
  \city{London}
  \country{United Kingdom}
}

\author{Emma Nasseri}
\email{enasseri@ripple.com}
\affiliation{%
  \institution{Ripple Labs, Inc.}
  \country{USA}
}

\author{Yebo Feng}
\email{yebo.feng@ntu.edu.sg}
\affiliation{%
  \institution{Nanyang Technological University}
  \city{Singapore}
  \country{Singapore}
}

\author{Lauren Weymouth}
\email{lweymouth@ripple.com}
\affiliation{%
  \institution{Ripple Labs, Inc.}
  \country{USA}
}

\renewcommand{\shortauthors}{Chen et al.}

\begin{abstract}
The rapid expansion of blockchain and digital asset ecosystems has intensified the challenge of translating academic research into deployable systems and regulatory frameworks. While advances in cryptography, consensus, digital assets, and governance are substantial, institutional mechanisms that sustain research-to-deployment translation at ecosystem scale remain comparatively under-theorized. This paper examines the architectural and coordination patterns that enable such translation, using the University Blockchain Research Initiative (UBRI) network as a representative case of long-term academic and industry collaboration. Drawing on research outputs and convenings from 2022 to 2025, we synthesize recurring design tensions across technical and institutional domains, including scalability versus security, decentralization versus governance, and privacy versus compliance. Rather than cataloging individual projects, we abstract system-level themes that connect research contributions to deployment constraints and policy adaptation, providing a structured lens for understanding how academic research informs production architectures, regulatory development, and ecosystem resilience in emerging decentralized infrastructures.
\end{abstract}

%%
%% CCS concepts — generate at https://dl.acm.org/ccs/ccs.cfm and paste here.
%%


\keywords{Ripple, University-Industry Collaboration, International Research Partnerships, Blockchain Innovation and Trends, Startup Ecosystem in Blockchain}

\maketitle

\section{Introduction}
\label{sect_introduction}

The expansion of blockchain and digital asset ecosystems over the past decade has revealed a persistent translation challenge: how can theoretical research be embedded into production systems and regulatory frameworks without compromising decentralization, security, or institutional legitimacy? As blockchain technologies move from experimental prototypes to large-scale financial and governance infrastructures, sustained collaboration between academia, industry, and policymakers becomes structurally necessary rather than optional. Yet the institutional architectures that sustain research-to-deployment translation at ecosystem scale remain comparatively under-theorized.

\begin{figure}[htbp]
\centering
\includegraphics[
  width=0.9\textwidth,
  height=0.95\textheight,
  keepaspectratio
]{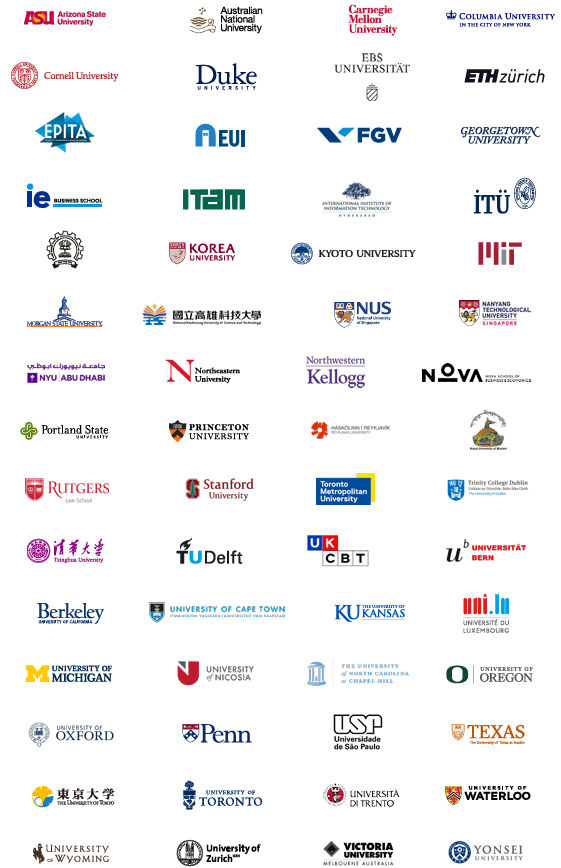}
\caption{Global distribution of UBRI research nodes, reflecting diverse institutional and regulatory environments across regions.}
\label{fig:ubri_partners}
\end{figure}

University-led research networks have emerged as one mechanism for addressing this constraint. Among these, the University Blockchain Research Initiative (UBRI), launched by Ripple in 2018, has developed a global network of over 60 academic partners spanning North America, Europe, Asia-Pacific, South America, and Africa (Fig.~\ref{fig:ubri_partners}). The network cuts across engineering, business, law, and policy, reflecting the interdisciplinary nature of blockchain system design. To date, the network has supported over 1,400 research projects, produced more than 1,200 peer-reviewed publications, and contributed to the development or revision of approximately 800 university-level courses, signaling sustained cross-disciplinary engagement at ecosystem scale. It has also established recurring convening mechanisms such as UBRI Connect and structured highlight reports to facilitate cross-sector knowledge exchange~\cite{feng2022university}.

The research emerging from this network spans multiple layers of the blockchain stack, including stablecoins and tokenized real-world assets, decentralized finance infrastructure, cryptographic security, oracle design, AI-assisted governance, interoperability, and broader institutional implications of blockchain adoption. Rather than presenting these efforts as isolated outputs, this paper organizes them into a systematized analysis of recurring technical and institutional tensions observed across sustained academic–industry collaboration.

Accordingly, this review synthesizes representative research directions from 2022 to 2025. We abstract recurring design tensions such as scalability versus security, decentralization versus governance, and privacy versus compliance, and examine how these tensions shape deployable architectures and ecosystem-level decisions. By framing UBRI as an illustrative case within the broader evolution of blockchain research translation, the paper provides a structured lens for understanding how academic inquiry informs production systems, regulatory development, and long-term ecosystem resilience.

\section{Research Themes and Design Patterns}
\label{sect_emerging_trends}
This section organizes UBRI-supported research by recurring technical and institutional
themes rather than by project or institution. The work spans cryptography, system
security, governance, interoperability, legal frameworks, and societal impact. Across
these areas, common design tensions recur: privacy versus compliance, decentralization
versus performance, automation versus accountability, and innovation versus regulation.
This section focuses on representative research that illustrates how academic work
informs system design, policy, and ecosystem outcomes.

To provide context for the research landscape, annual highlight reports synthesize
research contributions across multiple dimensions~\cite{ubrihighlight2025}. In the
remainder of this section, we organize representative works around recurring design
tensions to illustrate how academic insights inform system design, policy, and
ecosystem outcomes.

% Sect. 2.1. 
\subsection{Digital Money, Tokenization, and Regulatory Embedding Constraints}

Digital currencies and tokenized assets operate under a structural constraint: privacy, compliance, interoperability, and stability must be co-designed within monetary and asset infrastructures. The governing tension concerns how verification authority, disclosure granularity, and control rights are embedded across protocol and institutional layers rather than treated as separable policy objectives.

Selective disclosure architectures illustrate how privacy and compliance can be internalized within protocol design. Privacy classifications for offline CBDCs and compliance-compatible technical building blocks demonstrate how AML/CFT requirements can be embedded into system-level design parameters and transaction constraints~\cite{michalopoulos2025privacy}. Stablecoin systems further demonstrate that sanctions and KYC enforcement can coexist with confidentiality guarantees when privacy-preserving compliance mechanisms are embedded directly into distributed ledger architectures~\cite{duffie2025stablecoin}. Symbolic analysis of DeFi safety constraints clarifies guard conditions under oracle deviation scenarios~\cite{deng2024oracle}, while privacy-preserving identifier frameworks enable credential verification without exposing personally identifiable information~\cite{bradish2023covichain,kumari2025privacy}.

Public–private monetary coexistence and regulatory fragmentation further illustrate how governance structures and asset classification frameworks shape the evolution of digital monetary systems~\cite{guseva2025codependence,guseva2023fragmentation,guseva2024nftsecurities,werbach2024blockchain}. These institutional arrangements determine how control rights, compliance authority, and liability are distributed across public and private actors.

Cross-border interoperability and tokenization extend these constraints to systemic dependence. Foundational CBDC interoperability principles emphasize resilient cross-ledger architectures, while analyses of coupled financial and software networks demonstrate how interconnected systems can exhibit resilience–fragility trade-offs under stress~\cite{wef2023cbdc,fritz2024vulnerability,aymanns2023exit}. Simulation frameworks enable the evaluation of systemic coordination and liquidity dynamics in networked payment infrastructures~\cite{see2024pssimpy}, while legal analyses clarify interoperability requirements under data-protection and competition regimes~\cite{majcher2025interoperability}. Tokenized environmental assets further demonstrate the boundary between on-chain representation and off-chain enforcement~\cite{correia2025carbono}. Empirical evidence from digitally enforced collateral instruments shows how embedding enforcement rights into digital assets expands credit access in markets where traditional collateral mechanisms are unavailable~\cite{gertler2024digital}. Structural equilibrium analyses illustrate how governance design and regulatory coordination shape deployment outcomes~\cite{cao2024distributed}, and adaptive evaluation frameworks support systematic pre-deployment assessment~\cite{finan2024reinforcing}. Empirical analyses of CBDC adoption further identify trust, regulatory clarity, and perceived usefulness as primary determinants of user uptake, demonstrating that adoption barriers span technical, institutional, and behavioral dimensions~\cite{singh2025adoption}. Digital money and tokenization are thus bounded by regulatory embedding, governance structure, and systemic interdependence rather than by technical feasibility alone.

% sect 2.2
\subsection{Security and Adversarial Coordination Constraints}

Blockchain security is a coordination problem under adversarial conditions. The governing constraint concerns how ordering authority, censorship capability, vulnerability exposure, and governance response are distributed across protocol and institutional layers. Tensions therefore arise between performance and attack resistance, decentralization and enforceable guarantees, and transparency and defensive privacy. Security emerges not from isolated countermeasures, but from how control and incentives are structured throughout the system.

\subsubsection{Adversarial Control of Inclusion and Sequencing}

A central security boundary concerns transaction inclusion and ordering. Discretionary sequencing power creates structural tensions between validator incentives and user welfare, system efficiency and manipulation resistance, and decentralization and control over block production. The design problem is therefore not merely preventing isolated exploits, but constraining how ordering authority is allocated within consensus and execution layers.

Protocol-level approaches to transaction reordering mitigation embed ordering constraints directly into execution or consensus rules without relying on trusted intermediaries~\cite{alpos2023eating}. Formal analyses of block production behavior quantify how external pressures can influence selective transaction exclusion and alter neutrality assumptions in practice~\cite{wahrstatter2024blockchain}. Together, these strands show that resistance to MEV extraction and censorship depends on bounding discretionary inclusion power at the protocol layer. Sequencing authority becomes a governance boundary as much as a technical optimization problem.

\subsubsection{Detection, Adaptation, and Governance Response}

Security assurance further depends on how systems detect vulnerabilities, adapt to threats, and coordinate collective response. Precision analyses of low-level contract vulnerabilities, oracle integrity, and incentive-compatible verification mechanisms illustrate tradeoffs between detection coverage, accuracy, and operational cost~\cite{deng2024oracle}. Secure protocol designs for DeFi lending pools against oracle manipulation attacks demonstrate how vulnerability boundaries can be formalized and enforced directly within contract logic~\cite{arora2024secplf}. Decentralized oracle networks for verifiable counting extend incentive-compatible integrity guarantees to aggregation tasks beyond price feeds, addressing fault tolerance under adversarial conditions~\cite{nassirzadeh2024countchain}. Predictive models for identifying risky actors clarify statistical limits of automated screening under adversarial behavior~\cite{tsuchiya2024identifying}. Automated detection frameworks for price manipulation attacks in DeFi applications show how dynamic analysis and transaction pattern recognition identify adversarial behavior at the protocol layer~\cite{xie2024defort}. Benchmarking evaluations of large language models for code vulnerability detection further characterize how automated analysis scales to repository-level security assessment tasks~\cite{yildiz2025benchmarking}.

Beyond protocol-level vulnerabilities, systems must adapt to sophisticated crypto-enabled cybercrimes, particularly corporate-style ransomware operations. Recent empirical research suggests that, instead of relying on broad restrictions, effective governance responses can leverage blockchain transparency and digital footprint tracking to systematically monitor and disrupt these illicit networks ~\cite{cong2025anatomy}. Consequently, data-driven on-chain forensics serve as a key detection mechanism, reinforcing dynamic ecosystem security and coordinated responses under severe adversarial pressure.

Governance mechanisms mediate how decentralized systems respond to security incidents. Analyses of DeFi incentive structures and DAO governance practices show how institutional design shapes mitigation strategies and the perceived legitimacy of responses~\cite{kremer2024defi,werbach2024blockchain}. 
Human-in-the-loop verification frameworks demonstrate how incentive design can balance safety and liveness even under disagreement among participants~\cite{chen2022implementation}. 
Security, therefore, is not a static property but a dynamic equilibrium shaped by ordering rules, detection capacity, and governance coordination under pressure.

% Sect. 2.3.
\subsection{Consensus, Governance, and Market Coordination Constraints}

Protocol-level defenses against transaction reordering illustrate how ordering security can be embedded without reliance on centralized intermediaries~\cite{alpos2023eating}. 
Formal analyses of block production and censorship behavior quantify the systemic impact of regulatory or adversarial pressure on consensus outcomes~\cite{wahrstatter2024blockchain}. 
Game-theoretic analyses of Proof-of-Stake characterize validator equilibrium behavior under explicit reward–penalty structures~\cite{chen2024game}. 
Topology studies, robustness analyses of Federated Byzantine Agreement, and gossip-layer evaluations further demonstrate how communication structure, targeted attacks, and dissemination dynamics shape consensus resilience~\cite{tumas2023topology,tumas2023federated,scheidt2024gossipsub}. 
Controlled benchmarking frameworks extend this analysis by evaluating protocol robustness under simulated adversarial scenarios prior to deployment~\cite{touloupou2024validating}. System-level fuzz-testing frameworks complement controlled benchmarking by probing consensus algorithm behavior under automatically generated adversarial execution paths~\cite{kanhai2025rocket}.
Consensus security therefore emerges from the joint interaction of protocol design, incentive compatibility, and network structure.

Governance and market coordination extend these constraints to parameter control and value distribution. Learning-based governance frameworks illustrate adaptive parameter optimization under oversight constraints~\cite{xu2025autogov}. Privacy-preserving delegation protocols redefine transparency–accountability trade-offs in collective decision-making~\cite{nazirkhanova2025kite}. Analytical models of proposer–builder separation and liquidity provision reveal how incentive structures influence decentralization and efficiency outcomes~\cite{capponi2025liquidity,tang2025game}. Market impact analyses of large-order execution with endogenous liquidity supply further quantify how order size and strategic market-making jointly determine execution costs~\cite{capponi2025large}. Empirical work on the effects of futures market introduction on spot market quality illustrates cross-market information dynamics and their structural impact on price discovery and manipulation risk~\cite{augustin2023impact}. Interval-valued forecasting frameworks for cryptocurrency price ranges complement these analyses by characterizing market uncertainty through evolving computational models~\cite{maciel2023trading}. Comparative AMM implementations and integrated auction mechanisms further demonstrate how market architecture alters value redistribution~\cite{cruz2025xrplamm}. Shared-liquidity DEX designs for low-trading-volume tokens demonstrate how pool architecture reconfiguration sustains liquidity levels where standard market-making mechanisms are insufficient~\cite{singh2024deeper}. Governance and market structure are thus inseparable from consensus design; together they determine incentive alignment and systemic resilience. Broader surveys of blockchain and crypto economics synthesize these market, governance, and incentive strands into an integrated research agenda~\cite{biais2023advances}.

% Sect. 2.4.
\subsection{Privacy, Zero-Knowledge, and Cryptographic Constraints}

Cryptographic design defines the boundary between confidentiality and public verifiability in blockchain systems. The governing constraint is structural: correctness must remain publicly attestable while sensitive data remains hidden. This produces recurring tensions between proof efficiency and computational cost, anonymity and accountability, and present-day security assumptions versus post-quantum resilience.

Advances in zero-knowledge proof systems illustrate how verification can be achieved without exposing underlying data, decoupling computational correctness from information disclosure~\cite{chen2023hyperplonk}. Graph-based identity systems and privacy-preserving healthcare identifiers further demonstrate selective disclosure under accountability requirements~\cite{kumari2025privacy,bradish2023covichain}. Graph-based anonymous credential frameworks extend these mechanisms by constructing verifiable credentials from distributed identity graphs, enabling selective disclosure without exposing the underlying graph structure or membership~\cite{tang2024grac}. These developments collectively redefine admissible trust and verification boundaries in decentralized infrastructures.

Scalability and composability expose further limits. Partition-based privacy scaling approaches reorganize state to reduce storage overhead while preserving anonymity~\cite{cachin2025toxic}. SNARK-compatible verifiable encryption enables modular proof composition without sacrificing confidentiality~\cite{lee2024saver}. Privacy-preserving blockchain systems are therefore bounded by composability constraints across proofs, encryption, and identity layers.

% Sect. 2.5.
\subsection{Interoperability, Scaling, and System Constraints}

Blockchain scalability is not merely a performance problem; it defines the boundary conditions under which decentralization, interoperability, and security can coexist. The governing constraint concerns where computation, verification, and state transitions occur, and how trust assumptions are redistributed across layers. Tensions therefore arise between isolation and cross-network value transfer, on-chain execution and off-chain delegation, and efficiency gains versus preserved security margins.

Layered scaling architectures illustrate this redistribution of trust and verification authority. Trust-minimized optimistic rollup frameworks demonstrate how off-chain computation can retain cryptographic accountability through interactive fraud proofs while mitigating execution bottlenecks at the base layer~\cite{ye2024specular}. Payment channel networks further shift transaction flow off-chain while anchoring settlement security to base-layer dispute resolution guarantees~\cite{ersoy2022syncpcn}. In both cases, scaling reconfigures the locus of verification rather than eliminating it; security is inherited conditionally on fraud proofs, dispute windows, or settlement assumptions.

At the network layer, message propagation and market interaction dynamics impose additional constraints on system robustness. Improvements in propagation efficiency affect latency, synchronization, and fork dynamics~\cite{trestioreanu2024squelch}. Cross-market interaction analyses illustrate how information diffusion and financial signal integration influence system-level behavior~\cite{colombo2023interplay}. Optimization at this layer alters incentive alignment and coordination stability rather than simply increasing speed. Interoperability and scaling thus belong to a unified architectural question: how performance enhancement reshapes verification structure, trust assumptions, and systemic resilience.

% sect. 2.6.
\subsection{Sustainability and Social Impact Applications}

Blockchain-based sustainability initiatives expose a structural constraint: environmental and social objectives cannot be treated as external add-ons to financial infrastructure design. Instead, transparency, verification, incentive allocation, and energy consumption are co-determined at the architectural level. The governing tensions concern disclosure versus confidentiality, performance versus environmental cost, and innovation incentives versus distributive equity.

Green fintech research illustrates how financial and technological infrastructures embed sustainability trade-offs. Analyses of emerging green fintech agendas highlight concentration around blockchain-based verification mechanisms and the relative underdevelopment of alternative technological integrations~\cite{puschmann2024green}. Related work on climate impact in value chains emphasizes that environmental externalities propagate through governance structures, disclosure regimes, and incentive design~\cite{puschmann2023climate}. Systematic analyses of future financial system architectures examine how blockchain and digital technologies reshape financial infrastructure under evolving sustainability and regulatory constraints~\cite{puschmann2024financial}. Sector-specific applications in oil and gas demonstrate how blockchain-based transparency and supply chain provenance can be embedded in extractive industry operations to support environmental accountability~\cite{saraji2023sustainable}. Related work on renewable energy integration applies data science methods to characterize interdependencies among energy production, environmental outcomes, and economic growth, providing analytical foundations for blockchain-enabled sustainability monitoring~\cite{ikeda2024data}. Sustainability outcomes therefore depend less on isolated applications and more on how environmental constraints are internalized within system architecture.

% Sect. 2.7.
\subsection{Legal Frameworks and Regulatory Adaptation}

Decentralized execution does not eliminate jurisdictional authority; it redistributes regulatory attachment points across protocol layers, market interfaces, and dispute channels. The governing constraint is structural: global transaction reach operates within territorially bounded legal regimes, automated enforcement coexists with discretionary adjudication, and protocol autonomy remains embedded within regulatory oversight.

Research on decentralized markets analyzes regulatory fragmentation, digital asset classification, and evolving self-regulatory structures under distributed ledger infrastructures~\cite{guseva2024decentralized}. Parallel work on dispute resolution examines arbitration integration, digital evidence authentication, AI-assisted adjudication, and automated decision processes in transnational legal contexts~\cite{walters2025robots}. Comparative analyses of commercial and arbitration law across Asian, European, and North American jurisdictions extend this regulatory mapping to cross-border contexts, clarifying how enforcement authority, contract validity, and dispute resolution are constituted under digital economy conditions~\cite{walters2025commercial}. Research on the trajectory of AI governance further illustrates the co-evolution of regulatory instruments and technological capabilities, tracing how governance frameworks transition from voluntary principles toward binding regulatory structures as AI systems proliferate~\cite{chesterman2024evolution}. Across these strands, the architectural problem is not bypassing law, but designing systems that remain operationally decentralized while retaining enforceable pathways under legal plurality.

% Merged from sect. 2.8 and 2.9.
\subsection{Institutional Design, Social Impact, and Computational Paradigms}
Blockchain architectures implement institutional rules that determine incentive allocation, information visibility, and governance authority. The design space is defined by structural tensions: efficiency versus equity, protocol autonomy versus institutional embedding, automation versus oversight, privacy versus verifiability, and performance versus security margin. Architectural choices therefore propagate simultaneously across behavioral, market, and computational layers.

At the institutional layer, identity, market microstructure, and adoption dynamics are jointly shaped by how transparency, monitoring rights, and disclosure rules are encoded. System-dynamics modeling of identity adoption and breach externalities illustrates the co-evolution of incentives and security investment~\cite{mulaji2024dynamic}. Governance analyses of blockchain-based financial platforms further show how ledger architecture and incentive mechanisms reshape misreporting risks and coordination structures~\cite{cao2024distributed}. Empirical studies of financial-technology adoption indicate that sustained participation depends on the alignment between technical guarantees, trust formation, and user acceptance dynamics~\cite{puschmann2024fintech}. Empirical investigations of consumer resistance to crypto-payment identify trust deficits, perceived complexity, and regulatory uncertainty as structural barriers that impede adoption independently of technical readiness~\cite{sangari2024what}.

Computational constraints delimit feasible institutional configurations. Hybrid quantum-classical methods highlight transition challenges under evolving threat models~\cite{ghysels2024quantum}. Implementation-oriented research on quantum-safe ledgers demonstrates the practical feasibility of integrating post-quantum signature schemes into existing blockchain architectures, using the XRPL as a concrete case study~\cite{lele2025quantum}. Optimization of proof systems and authenticated data structures reduces verification overhead while preserving cryptographic integrity~\cite{luo2023fast,choi2024lmpt}. Privacy-preserving ownership transfer protocols further demonstrate how zero-knowledge constructions can maintain structured disclosure under decentralized coordination~\cite{badakhshan2023privacy}. Computational enhancement thus modifies admissible governance and privacy configurations rather than merely improving throughput.

\section{Ecosystem Expansion Beyond Academia}
\label{sect_ecosystem_expansion}
The expansion of blockchain research beyond academia is shaped by a research-to-deployment coordination challenge. Theoretical contributions must be embedded within regulatory, institutional, and operational infrastructures without undermining research independence. Such translation requires the reconfiguration of authority, accountability, and coordination structures across academic and non-academic domains. The central challenge extends beyond technical deployment. It concerns sustaining neutrality, credibility, and technical rigor while engaging with real-world systems influenced by political, financial, and market dynamics.

\subsection{Institutional and Policy Bridging Mechanisms}

At the governance mediation layer, research-to-implementation translation exhibits recurring institutional and architectural patterns. Privacy–compliance systems illustrate selective disclosure as a reusable design logic across identity, healthcare, and regulated digital infrastructures. Infrastructure–application distinctions similarly expose friction: protocol-level work requires long-term operational stability and backward compatibility, whereas application development prioritizes usability, offline resilience, and domain-specific constraints. In practice, these differences affect funding cycles, deployment timelines, and evaluation criteria.

\begin{figure}[htbp]
\centering
\includegraphics[width=0.9\textwidth]{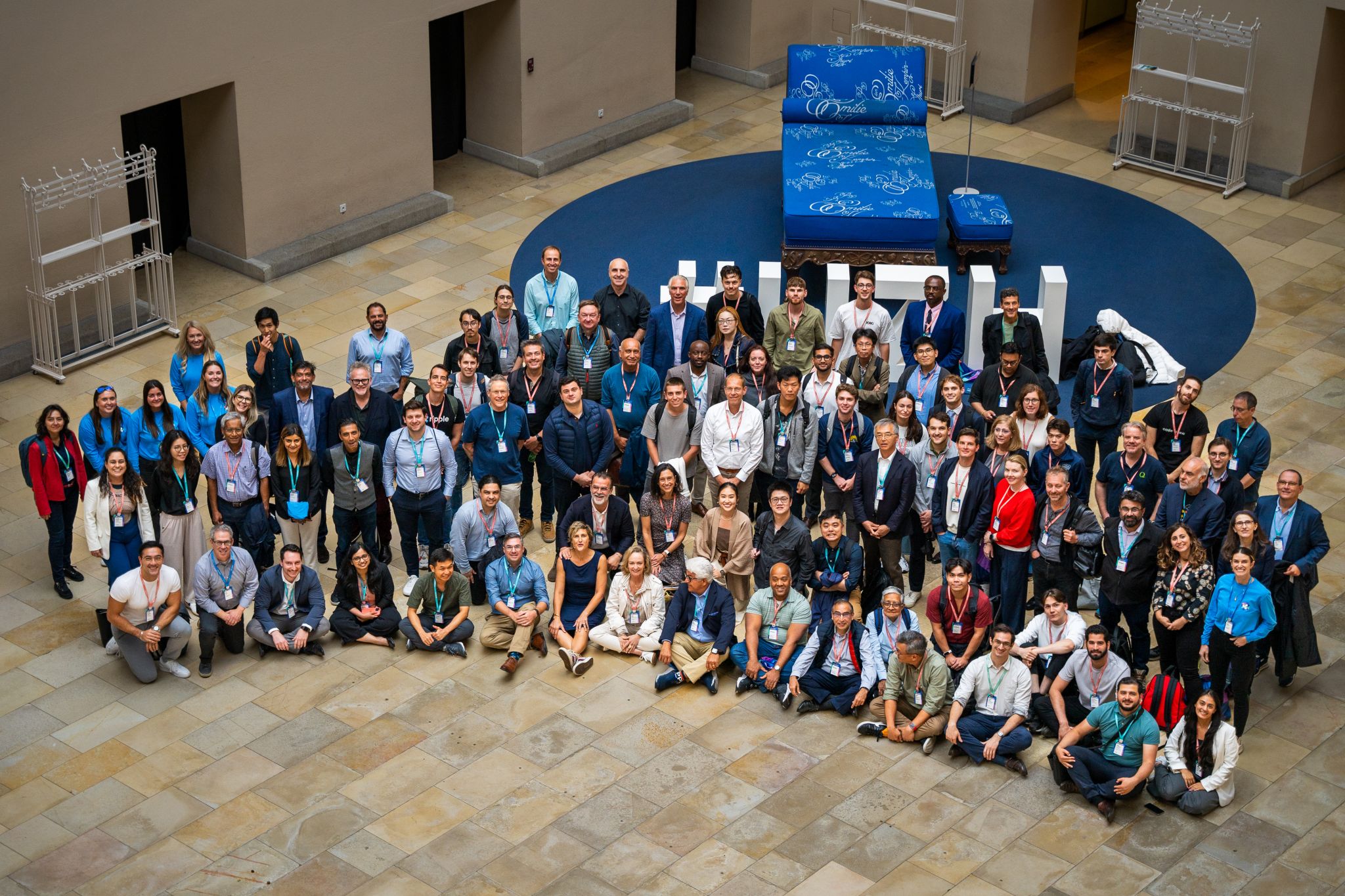}
\caption{UBRI Connect 2024}
\label{fig:ubri_connect_2024}
\end{figure}

Policy engagement demonstrates the importance of institutionalized mediation spaces rather than episodic outreach. One-time symposia can facilitate direct discussion of pending legislation, but sustained event series build regulatory literacy and long-term trust among stakeholders who otherwise lack neutral venues for dialogue. Location choices, capital cities for legislative access, financial centers for supervisory agencies, or regional hubs for emerging ecosystems, reflect strategic decisions about which actors can be effectively convened. Figure~\ref{fig:ubri_connect_2024} illustrates such a neutral convening architecture.

\begin{figure}[htbp]
\centering
\includegraphics[width=0.9\textwidth]{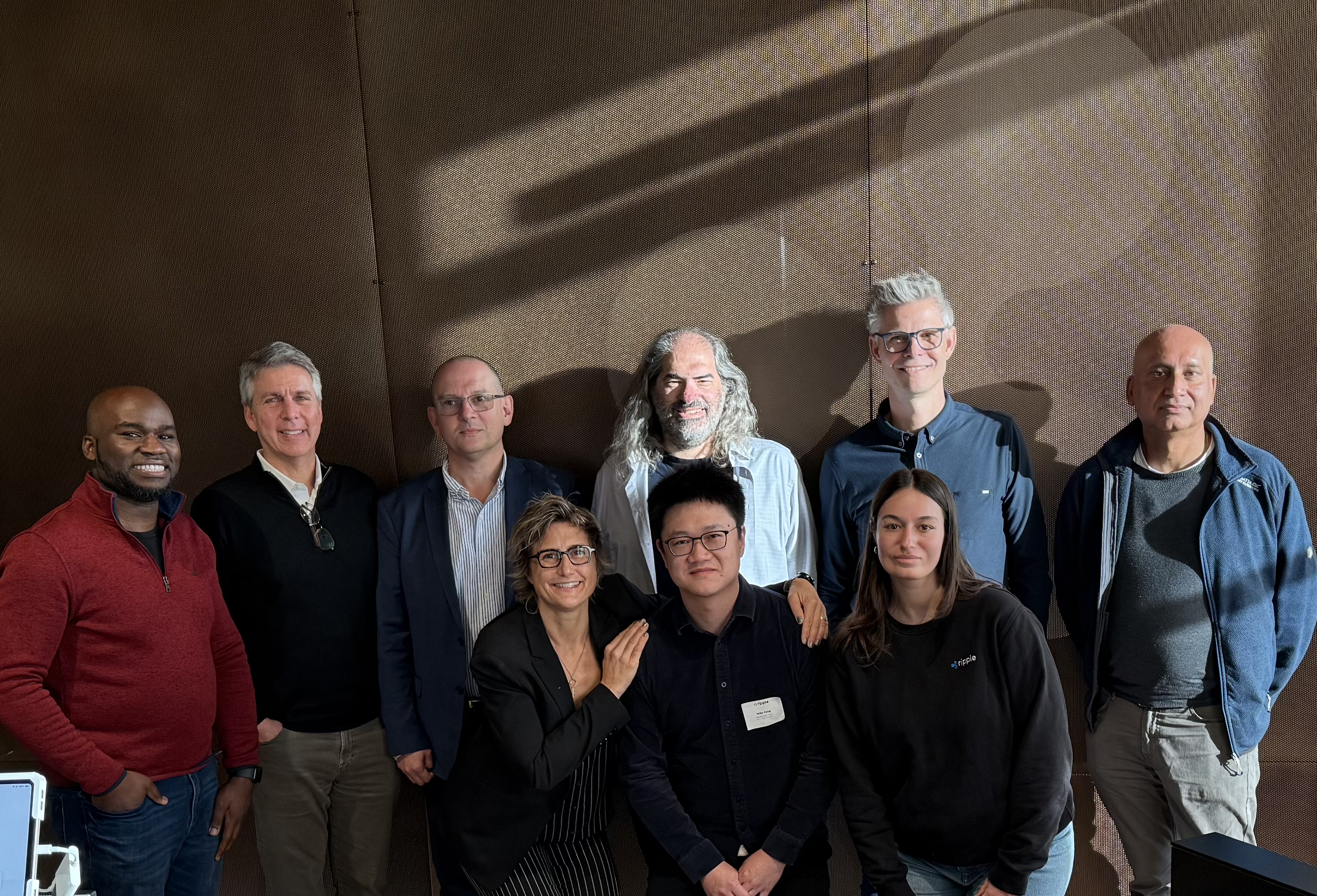}
\caption{An advisory council session illustrating cross-sector engagement between industry practitioners and academic researchers within the UBRI network.}
\label{fig:advisory_council}
\end{figure}

Formal advisory structures further redistribute technical influence while preserving autonomy. The design tension lies between breadth and depth: broad representation ensures cross-domain insight, yet smaller councils enable actionable technical coordination. Figure~\ref{fig:advisory_council} reflects this balance between inclusiveness and operational effectiveness.

\subsection{Innovation Incentives and Developer Pipeline Architectures}

At the innovation acceleration layer, event architectures, conferences, competitions, hackathons, and accelerators serve as coordination infrastructures that compress research-to-implementation cycles. Their structure shapes outcomes. Competitions with narrowly defined problem domains often generate technically deep but domain-constrained prototypes, whereas broader challenge formats encourage experimentation but risk diffusion of effort. Similarly, hackathons emphasize rapid prototyping and experiential learning, while accelerator programs prioritize venture sustainability and ecosystem integration.

Funding models reveal further tradeoffs. Equity-based accelerators align long-term financial incentives but may constrain academic neutrality; non-equity grant models preserve research independence yet require alternative sustainability mechanisms. Platform-specific programs enable deep technical integration at the cost of cross-platform experimentation. Representative implementations illustrate how decentralized identity wallets, verifiable credential systems, and tokenized asset frameworks migrate from research prototypes into user-facing applications, embedding architectural principles into production environments.

\begin{figure}[t]
\centering
\includegraphics[width=0.8\textwidth]{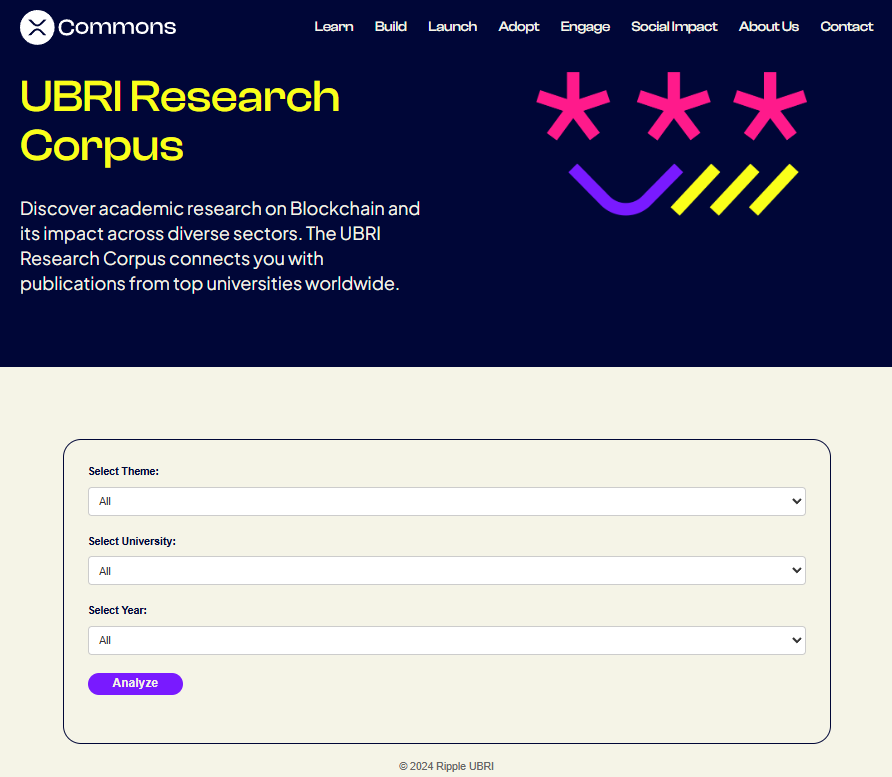}
\caption{UBRI Research Corpus - A comprehensive database for discovering blockchain research from UBRI partner institutions~\cite{ubriresearchcorpus2024}}
\label{fig:research_corpus}
\end{figure}

Research discovery infrastructure supports this translation layer. The UBRI Research Corpus provides centralized access to distributed academic outputs, enabling thematic filtering and cross-institutional discovery~\cite{ubriresearchcorpus2024}. Figure~\ref{fig:research_corpus} illustrates this integration layer, which reduces information fragmentation and lowers barriers between academic knowledge and practitioner needs.

\subsection{Experiential Learning and Capacity Formation}

At the capability formation layer, experiential programs address the ecosystem-level constraint of talent formation. Time-compressed hackathon formats prioritize rapid experimentation and exposure to platform tooling, while multi-week residency models enable deeper architectural engagement and iterative refinement. The XRPL Student Builder Residency exemplifies a structured mentorship model in which theoretical instruction, peer review, and prototype deployment occur in sequence. As summarized in Table~\ref{tab:residency_outcomes}, measurable skill gains differ across cohorts, reflecting how participant background and program intensity shape learning trajectories.

\begin{table}[t] 
\caption{XRPL Student Builder Residency Learning Outcomes}
\label{tab:residency_outcomes}
\centering
\begin{tabular}{lcc}
\hline \hline
\textbf{Metric} & \textbf{Cohort 1.0} & \textbf{Cohort 2.0} \\
\hline
Term & Fall 2024 & Summer 2025 \\
Class Size & 12 & 17 \\
Participant Profile & Novice & Intermediate/Advanced \\
\hline
\multicolumn{3}{c}{\textbf{Average Skill Increase}} \\
\hline
Development Skills & 87\% & 49\% \\
General Blockchain Knowledge & 117\% & 62\% \\
XRPL-Specific Knowledge & 219\% & 153\% \\
\hline \hline
\end{tabular}
\end{table}

\begin{figure}[htbp]
\centering
\subfloat[Workshop phase\label{fig:residency_a}]{%
  \includegraphics[width=0.48\textwidth]{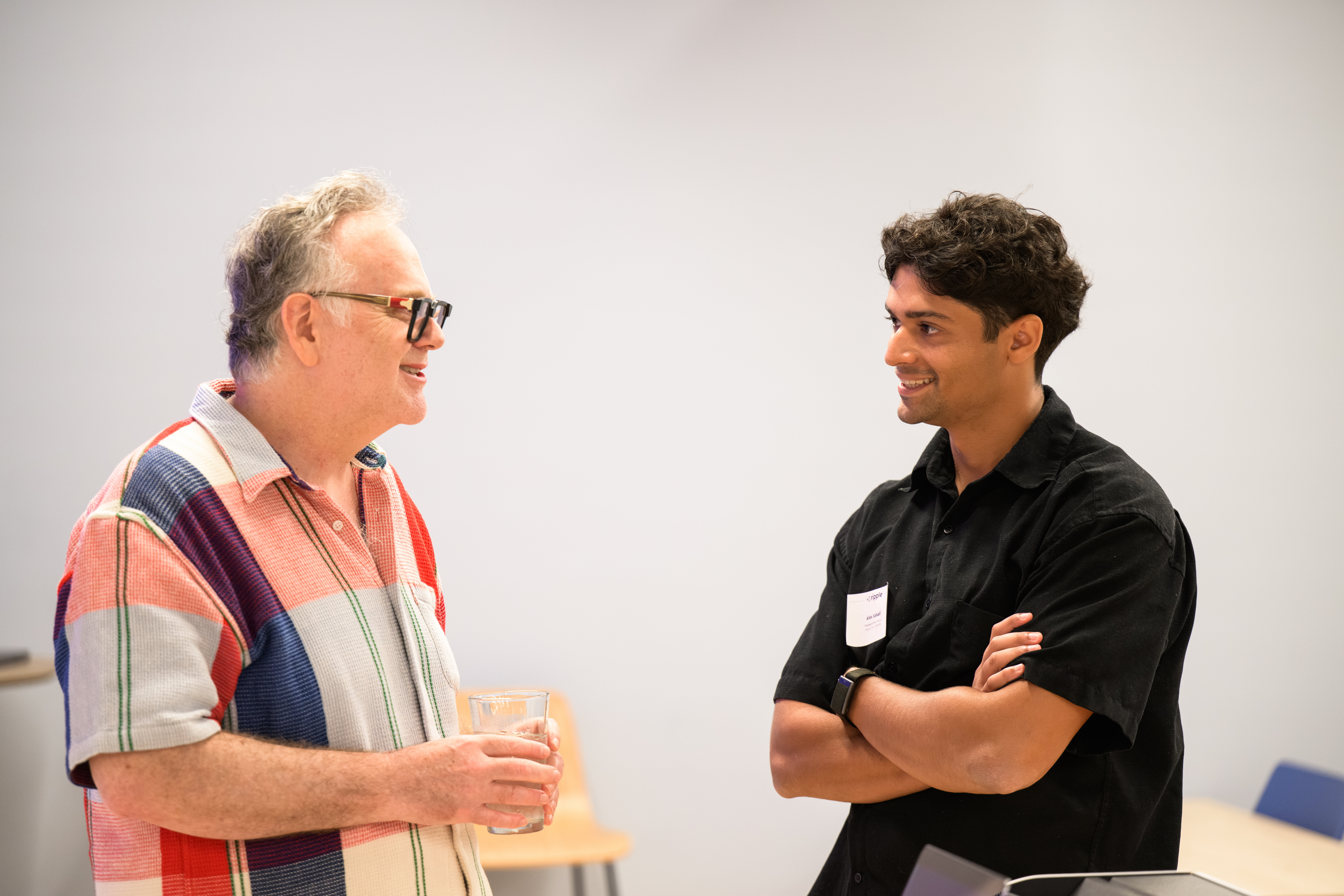}}
\hfill
\subfloat[Demo day\label{fig:residency_b}]{%
  \includegraphics[width=0.48\textwidth]{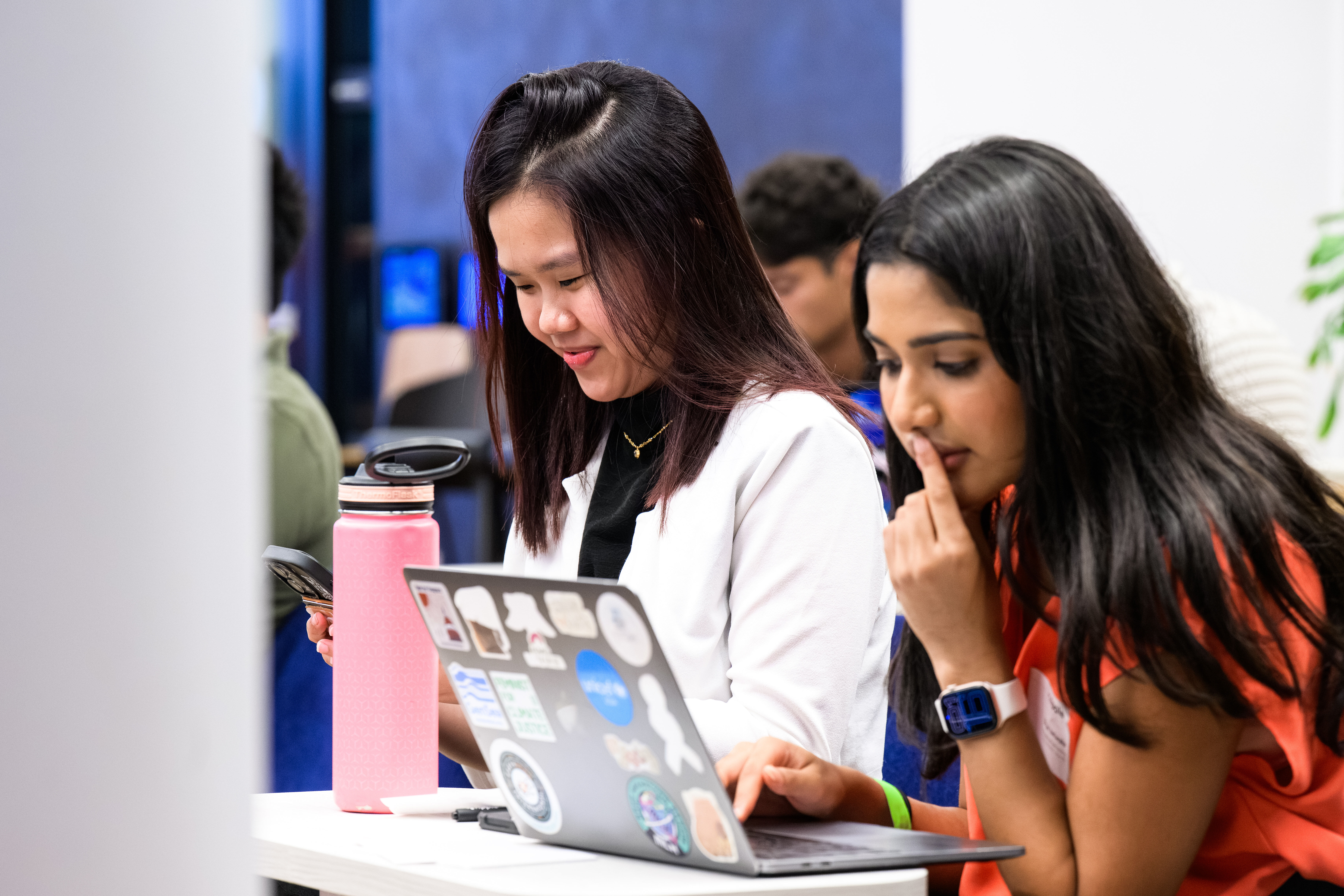}}
\caption{The XRPL Student Builder Residency: From architectural design to project prototype.}
\label{fig:residency_combined}
\end{figure}

Across these layers, ecosystem expansion follows a recurring architectural logic. Research translation stabilizes when governance mediation, innovation acceleration, and capability formation are jointly configured rather than treated as isolated interventions. Academic neutrality functions not as isolation from practice, but as a structural condition that enables sustained multi-stakeholder engagement while preserving analytical rigor.

\section{Open Questions and Research Directions}
\label{sect_limitation}

The systematization presented in this review highlights several areas where further research remains necessary to deepen understanding of research-to-deployment translation in blockchain ecosystems. While the analysis draws on a substantial body of work supported within the UBRI network between 2022 and 2025, it should be understood as an illustrative institutional case rather than an exhaustive mapping of the global blockchain research landscape. Future studies may extend this framework by incorporating comparative analyses across other research consortia, industry-led initiatives, and regulatory environments.

A second open direction concerns the relative scarcity of longitudinal empirical evidence. Much of the reviewed literature focuses on protocol design, incentive mechanisms, and theoretical tradeoffs. However, systematic measurement of long-term adoption dynamics, user behavior, governance performance, and institutional adaptation remains limited. Expanding empirical and data-driven inquiry will be essential for evaluating how blockchain systems function once deployed at scale and under evolving regulatory conditions.

Finally, several structural questions remain unresolved. How do incentive mechanisms evolve under sustained participation and changing market environments? How do governance frameworks adapt over extended time horizons, particularly in hybrid systems that combine decentralized coordination with regulatory oversight? To what extent do legal and institutional contexts shape system outcomes in practice? Addressing these questions will require deeper collaboration across technical, economic, and legal disciplines, as well as tighter integration between empirical measurement and system-level design analysis.

By articulating these open directions, this review seeks not to delimit the field but to provide a structured foundation for future inquiry into the institutional maturation of decentralized infrastructures.
\section{Conclusion}
\label{sect_conclusion_outlook}

This review has examined the institutional architectures that sustain research to deployment translation in blockchain ecosystems, using the UBRI network as an illustrative case of long-term industry and academia coordination. Rather than evaluating individual project outcomes, the analysis has focused on recurring structural patterns through which academic research interacts with regulatory frameworks, operational infrastructures, and market environments.

Across technical domains, including stablecoins and tokenized assets, smart contract security, AI-assisted governance, privacy-preserving cryptography, and interoperability, common design tensions persist. Scalability remains coupled with decentralization tradeoffs. Privacy mechanisms must negotiate compliance constraints. Governance automation introduces accountability challenges. Cross-chain interoperability raises questions of jurisdictional and institutional alignment. These tensions are not isolated technical problems but system-level coordination constraints that shape how research advances migrate into deployable infrastructures.

The UBRI case illustrates how sustained funding structures, recurring convening mechanisms, research discovery infrastructures, and cross-sector advisory processes can stabilize this translation layer without fully collapsing academic independence into industry imperatives. In this sense, the contribution of the network lies less in any individual innovation than in demonstrating how institutional design mediates the relationship between theoretical inquiry and real-world deployment.

Looking ahead, the continued evolution of blockchain ecosystems will depend not only on technical breakthroughs but also on governance architectures capable of aligning incentive structures, regulatory adaptation, and interdisciplinary collaboration at scale. Future research should further examine how translation infrastructures can preserve neutrality, foster long-term resilience, and adapt to rapidly shifting legal and technological environments. By abstracting institutional patterns from empirical cases, this review contributes a framework for understanding how decentralized infrastructures can mature without sacrificing research integrity or systemic stability.

\begin{acks}
The authors thank Dr.\ Campbell R. Harvey and Dr.\ Lin William Cong for their support, revision, and guidance.
The authors also thank the UBRI partner institutions and researchers whose work is discussed in this paper.
This research was supported by the University Blockchain Research Initiative (UBRI), sponsored by Ripple Labs, Inc.
\end{acks}

\section*{Author Contributions}

CCC and YW jointly developed the conceptual framework, conducted the literature synthesis, and wrote the manuscript. EN contributed Section~3 on ecosystem expansion. LW and YF reviewed and revised the manuscript.

\section*{Conflict of Interest}

E. Nasseri and L. Weymouth are employees of Ripple Labs, Inc.
C.-C. Chen is a part-time affiliate of Ripple Labs, Inc.
Y. Feng serves as a member of an advisory council associated with Ripple Labs, Inc.
The remaining author declares no known conflicts of interest.

\bibliographystyle{ACM-Reference-Format}
\bibliography{reference}

\end{document}